\documentclass[pra,showpacs,aps,amsmath,amssymb,twocolumn]{revtex4-1}
\usepackage{graphicx}
\usepackage{graphics}
\usepackage{amsmath,amsfonts,amssymb}
\usepackage{bm}
\usepackage{epsfig}
\usepackage[english,french]{babel}
\usepackage[utf8]{inputenc}
\usepackage{xcolor}
\usepackage{ulem}

\begin{document}
\title{Modelling laser-atom interactions in the strong field regime}
\author{A. Galstyan$^{1}$,
Yu. V. Popov$^{2,3}$, F. Mota-Furtado$^{4}$, P.F. O'Mahony$^{4}$, N. Janssens$^{1 }$, S.D. Jenkins$^{4}$, O. Chuluunbaatar$^{3,5}$ and B. Piraux$^{1}$
}                     

\affiliation{$^1$ Institute of Condensed Matter and Nanosciences, Universit\'e Catholique de Louvain, 2 chemin du cyclotron,\\
                 $^{\;\;\;}$Box L7.01.07, B-1348 Louvain-la-Neuve, Belgium. \\
$^2$ Skobeltsyn Institute of Nuclear Physics, Lomonosov Moscow State University, Moscow, Russia.\\
$^3$ Joint Institute for Nuclear Research, Dubna, Russia.\\
$^4$ Department of Mathematics, Royal Holloway, University of London, Egham, Surrey TW20 0EX, United Kingdom.\\
$^5$ Institute of Mathematics and Computer Science, National University of Mongolia, UlaanBaatar, Mongolia.}
\begin{abstract}
We consider the ionisation of atomic hydrogen by a strong infrared field. We extend and study in more depth an existing semi-analytical model. Starting from the time-dependent Schr\"odinger equation in momentum space and in the velocity gauge we substitute the kernel of the non-local Coulomb potential by a  sum of $N$ separable potentials, each of them supporting one hydrogen bound state.  This leads to a set of $N$ coupled one-dimensional linear Volterra integral equations to solve. We analyze the gauge problem for the model, the different ways of generating the separable potentials and establish a clear link with the strong field approximation which turns out to be a limiting case of the present model. We calculate electron energy spectra as well as the time evolution of electron wave packets in momentum space. We compare and discuss the results obtained with the model and with the strong field approximation and examine in this context, the role of  excited states. 
\end{abstract}
\maketitle
\section{Introduction}
Numerous experiments have been carried out to study the highly non-linear interaction of a one-active electron system with an intense infrared laser pulse. Such experimental studies  have led to the development of various mathematical methods and numerical algorithms to solve the corresponding Time-Dependent Schr\"odinger Equation (TDSE). Although the numerical solution of the TDSE has provided significant insights into the basic processes resulting from this interaction, namely Above-Threshold Ionisation (ATI) and High-Order Harmonic Generation (HOHG), the comparison of the theoretical results with the experimental data is made difficult because in the experiments, a precise characterization of the temporal and spatial distributions of the laser beam is in general not possible. Furthermore, it is usually very difficult to draw conclusions regarding the actual physical mechanism underlying these processes since it is always {\it after} the laser turnoff that the relevant information about various observables is extracted from the solution of the TDSE.\\

In order to extract valuable information about the interaction mechanisms, one has to rely on models. The first one that laid the foundation of our understanding of laser-atom interactions is due to Keldysh. In his seminal 1965 paper \cite{Keldysh}, he describes, in the length gauge, the ionisation of an atom as the result of the transition from its ground state to a dressed continuum state. This dressed continuum is a Volkov state \cite{Volkov}. It takes into account the dipole interaction potential at all orders while neglecting completely the binding potential. It is important to note that in this model, the role of all the excited states of the atom is neglected. In addition, Keldysh introduced a so-called adiabaticity parameter $\gamma=\omega\sqrt{2I_p}/E$ where, in atomic units, $\omega$ is the photon energy, $I_p$ the ionisation potential and E the electric field amplitude. He showed that in the limit $\gamma\ll 1$, tunneling is the dominant ionisation process while for $\gamma\gg 1$, ionisation results from multiphoton transitions. Later, Keldysh theory has been generalized by Faisal \cite{Faisal} and Reiss \cite{Reiss} who developed an approach based on the S-matrix formalism in the velocity gauge. It takes into account high-order terms describing multiple rescatterings of the so-called indirect electrons by the Coulomb potential of the residual ion. By contrast to the direct electrons that escape without returning to the residual ion, indirect electrons undergo a quiver motion driven by the oscillating field. The S-matrix treatment, which can be developed also in the length gauge, reduces to Keldysh's result at the first order. It is after Faisal and Reiss papers that the terminology "Strong Field Approximation" (SFA) appeared with  the underlying theoretical treatment usually referred to as the Keldysh-Faisal-Reiss (KFR) theory.\\

It is important to mention that KFR theory is not the only one that leads to the theoretical framework of the SFA. Perelomov, Popov and Terent'ev  (PPT) \cite{Per1,Per2,Per3} arrived at  expressions similar to those obtained within the SFA by using a formalism based on a time-dependent Lippmann-Schwinger like equation. Recently, we reformulated the SFA and showed by introducing anzatses that all SFA theoretical schemes may be grouped into a set of families of schemes \cite{Galstyan}.\\

Despite various inherent theoretical problems such as the gauge, the SFA theoretical schemes have provided analytical results that have helped to interpret qualitatively the numerical solution of the TDSE  and to extract the underlying mechanisms. Furthermore, in SFA based approaches, the analytical expressions for various observables involve the classical action thereby establishing a bridge between quantum and classical calculations. However, it is important to stress that in the SFA approaches, the infinite range of the Coulomb potential is completely neglected or, equivalently, all the excited states of the atom are not taken into account. So, strictly speaking, the SFA approaches apply to negative ions only, even though a lot of effort has been made to introduce corrections that take account of the long range of the Coulomb potential through, for instance, the classical action \cite{Per3}.\\

 In order to study in detail the role of the long range of the Coulomb potential in the interaction of atomic hydrogen with a strong laser field in a regime where tunnel ionisation dominates ($\gamma\ll 1$), we have developed more in depth in this paper a model which has been introduced by Tetchou {\it et al.} \cite{Tetchou}. In this model, one starts with the TDSE in momentum space, uses the velocity gauge and replaces the kernel of the nonlocal Coulomb potential by a sum of $N$ separable potentials, each of them supporting one bound state of atomic hydrogen. Under these conditions, the TDSE reduces to a set of $N$ coupled one-dimensional linear integral Volterra equations of the second kind.  \\

This model which goes far beyond the SFA, has undeniable advantages. The first one is its computational simplicity allowing, for instance, to explore the very long wavelength limit without difficulty.  With our model including only the 1s state of atomic hydrogen, we are pre\-sently working at a wavelength of 40 $\mu\mathrm{m}$. The second important advantage is the fact that it can be generalized, within the mean field approximation to the treatment of one-electron processes in complex systems like a heavy molecule exposed to an ultrashort laser pulse.\\

In this contribution, we treat several key aspects of the model namely  the generation of the separable potentials, the gauge problem and the link with the SFA. We then calculate electron energy spectra and the time evolution of electron wave packets in momentum space. The study of the time evolution  of the electron wave packet in momentum state provides  direct information on the ionisation time as well as on the ionisation mechanism when the Keldysh parameter $\gamma\ll 1$. In addition, by comparing our results obtained with our model and within the SFA, we can assess the pertinence of the predictions of the SFA regarding the ionisation mechanism. \\

The paper is divided in three main sections in addition to this introduction. The second section is devoted to the theory. We  describe the key equations of the model and discuss methods to generate the separable potentials, the gauge problem and the connection with the SFA. In section $3$, we present and discuss our results in detail. The last section is devoted to the conclusions and perspectives. Atomic units combined with the gaussian system for the electromagnetic field are used through unless otherwise specified.

\section{Theory}

\subsection{Preliminary remark}
In order to study the role of the Coulomb potential when atomic hydrogen interacts with a strong oscillating electric field in the tunneling regime, it is more natural to work in the velocity gauge and in  momentum space \cite{deBohan}. This is a consequence of Ehrenfest's theorem:
\begin{equation}
\frac{\mathrm{d}}{\mathrm{d}t}<\vec{p}>=-\langle\nabla V\rangle,
\end{equation}
where $\vec{p}$ is the canonical momentum and $V$ the Coulomb potential. This equation means that the time evolution of the canonical momentum results only from the gradient of the Coulomb potential. Therefore, in the limit $\langle\nabla V\rangle\rightarrow 0$, the canonical momentum becomes a constant of motion which reduces to the drift velocity (in a.u.) of the ejected electron. This contrasts with the length gauge where, from the point of view of the forces acting on the electron, the role of the Coulomb potential is decoupled from the electric field. \\

For the sake of clarity, let us recall that in the confi\-guration space and irrespective of the gauge, the canonical momentum is defined by:
\begin{equation*}
\vec{p}=-i\vec{\nabla}_{\vec{r}}.
\end{equation*}
In the length gauge, this canonical momentum reduces to:
\begin{equation*}
\vec{p}=\vec{\pi},
\end{equation*}
where the mechanical momentum $\vec{\pi}=m\dot{\vec{r}}$ with $\vec{r}$ the position vector of the electron. In the velocity gauge, the canonical momentum writes:
\begin{equation*}
\vec{p}=\vec{\pi}-\frac{1}{c}\vec{A}(t),
\end{equation*}
where $\vec{A}(t)$ is the potential vector and $c$, the speed of light.

\subsection{Basic equations}
We start from the TDSE which governs the dynamics of  atomic hydrogen exposed to an external oscillating field. We work in momentum space and use the velocity gauge within the dipole approximation. For the sake of simpli\-city, we assume that the electric field is linearly polarized along the unit vector $\vec{e}$ the direction of which coincides with the z-axis.  Note that any other type of polarization can be treated without additional difficulty. In these conditions, the TDSE is:\\
\begin{eqnarray*}
\left [\mathrm{i}\frac{\partial}{\partial t}-\frac{p^2}{2}-\frac1c A(t)(\vec e\cdot\vec p)-
\frac{1}{2c^2}A^2(t)\right]\Psi(\vec p,t) + 
\end{eqnarray*}
\begin{equation}
+\int\frac{\mathrm{d}\vec{u}}{(2\pi)^3}V(\vec p-\vec u)\Psi(\vec u,t)=0,\;\;\;\;\Psi(\vec{p},0)=\varphi_{1s}(\vec{p}).
\end{equation}
$V(\vec{p}-\vec{u})$ is the kernel of the Coulomb potential. The main idea of the model is to 
replace this kernel by a sum of $N$ symmetric separable potentials supporting $N$ bound states of atomic hydrogen: 
\begin{equation}
V(\vec p-\vec u)=\frac{4\pi}{|\vec p-\vec u|^2}\approx V(\vec{p},\vec{u})=\sum_{n=1}^Nv_n(\vec p)v_n^*(\vec u)
\end{equation}
Before looking for the solution of Eq. (2) with the kernel replaced by the finite sum of separable potentials, it is convenient to define the following quantities involving the potential vector:
\begin{eqnarray}
\zeta(t)& = &\frac{1}{2c^2}\int_0^t[A(\xi)]^2\mathrm{d}\xi,\\
b(t)&=& - \frac1c \int_0^t A(\xi)\mathrm{d}\xi,\\
b'(t) &=& \frac{1}{\omega}\sqrt{\frac{I}{I_0}}\sin^2\left[\frac{\pi t}{T}\right]\sin(\omega t).
\end{eqnarray}
Eqs (5) and (6) and the fact that $A(t)/c=-b'(t)$ define the potential vector as a sine square pulse of frequency $\omega$ and peak intensity $I$ in W/cm$^2$.
$I_0=3.5\times 10^{16}$ W/cm$^2$ is the atomic unit of intensity and the total duration of the pulse is $T=2\pi N_c/\omega$ where $N_c$ is the number of optical cycles within the pulse. In all the results presented here, we used a sine square pulse but any other  pulse shape may be considered.\\

In order to solve Eq. (2) with the kernel of the Coulomb potential given by Eq. (3), we first perform a contact transformation of the wave function $\Psi(\vec{p},t)$ to eliminate the $A^2(t)$ term from Eq. (2),
\begin{equation}
\Psi(\vec{p},t)=e^{-\mathrm{i}\zeta(t)}\Phi(\vec{p},t),
\end{equation}
and then define the following function:
\begin{equation}
F_n(t)=\int\frac{\mathrm{d}\vec{u}}{(2\pi)^3}v^*_n(\vec u)\Phi(\vec u,t).
\end{equation}
Under these conditions, the solution of Eq. (2) in which the kernel of the Coulomb potential is replaced by a sum of $N$ symmetric separable potentials may be written formally as follows:
\begin{eqnarray*}
\Phi(\vec p,t)=e^{-\mathrm{i}\frac{p^2}{2}t+\mathrm{i}b(t)p_z}\left[\Phi(\vec p,0)+\mathrm{i}\sum_{n=1}^Nv_n(\vec p)\times\right.
\end{eqnarray*}
\begin{equation}
\left.\times\int_0^t\mathrm{d}\xi F_n(\xi)e^{\mathrm{i}\frac{p^2}{2}\xi-\mathrm{i}b(\xi)p_z}\right].
\end{equation}
In order to find the unknown functions $F_n(\xi)$  we substitute $\Phi(\vec{u},t)$ in Eq. (8) by the expression (9) given above. We obtain a system of $N$ coupled linear Volterra integral equations of the second kind which may be written in matrix form as :

\begin{equation}
\mathbf{F}(t)=\mathbf{F}_0(t)+\int_0^t\mathbf{K}(t,\xi)\mathbf{F}(\xi)\mathrm{d}\xi.
\end{equation}
$\mathbf{F}(t)$ is a vector of dimension $N$, the components of which are the  $F_n(t)$ functions. $\mathbf{F}_0(t)$ is a vector of the same dimension  which comes from the contribution of the initial state $\Phi(\vec{p},0)$ present in Eq. (9). $\mathbf{K}(t,\xi)$ which is the kernel of the Volterra integral equation is a $N\times N$ matrix (see \cite{Tetchou} for the details of the calculations). Consequently, we have reduced the 4-dimentional TDSE to a system of $N$ 1-dimentional Volterra integral equations. 

\subsection{Generation of the separable potentials}

Depending on the constraints we want to impose, there are two different methods to generate the separable potentials. The first method consists in imposing that the sum of the $N$ symmetric separable potentials supports $N$ bound states of atomic hydrogen. In momentum space, the exact wave function $\varphi_j(\vec{p})$ associated to the $j^{th}$ bound state of energy $\varepsilon_j$, satisfies the stationary Schr\"odinger equation:
\begin{equation}
\left (\varepsilon_j-\frac{p^2}{2}\right)\varphi_j(\vec p)+\sum_{n=1}^N 
\left[\int\frac{\mathrm{d}\vec{u}}{(2\pi)^3}v^*_n(\vec u)\varphi_j(\vec u)\right]v_n(\vec p)=0,
\end{equation}
Since this equation must be satisfied for $j=1,...,N$, we can rewrite it in matrix form:
\begin{equation}
\mathbf{\Phi+AV}=0,
\end{equation}
where $\mathbf{\Phi}$ is a vector with $N$ components given by the first term of the lhs of Eq. (11). The components of vector $\mathbf{V}$  are the unknown functions $v_n(\vec{p})$. $\mathbf{A}$ is a $N\times N$ block-diagonal matrix, each block corresponding to a given value of the angular momentum $\ell$. The elements of matrix $\mathbf{A}$ are given by the term between square brackets in the lhs of Eq. (11). Provided that $\mathbf{A}^{-1}$ exists,  we have:
\begin{equation}
\mathbf{V}=-\mathbf{A}^{-1}\mathbf{\Phi}.
\end{equation}
As such, this equation cannot be used since the elements of matrix $\mathbf{A}$ depend on the unknown functions $v_n(\vec{p})$. In order to solve this problem, we introduce the symmetric $N\times N$ matrix $\mathbf{\Gamma}$ defined by:
\begin{equation}
\mathbf{\Gamma}=\mathbf{AA}^T,
\end{equation}
where $\mathbf{A}^T$ denotes the transpose matrix of $\mathbf{A}$. The elements of  the matrix $\mathbf{\Gamma}$ are known and given by:
\begin{equation}
\mathbf{\Gamma}_{ij}=\int\frac{\mathrm{d}\mathbf{p}}{(2\pi)^3}\varphi^*(\mathbf{p})\left(\frac{1}{2}p^2-\varepsilon_j\right)\varphi_j(\mathbf{p}).
\end{equation}
Eq. (14) has in fact a finite number of solutions for the ele\-ments of matrix $\mathbf{A}$ in which we omit a global phase factor. It is therefore necessary to impose some prescription. We could , for instance, assume that any $\ell$-block of $\mathbf{A}^{-1}$ is triangular or symmetric. In fact we have shown that both prescriptions give results that are not significantly different. All the results presented here have been obtained by assuming that all the $\ell$-blocks of $\mathbf{A}^{-1}$ are triangular.\\

 For the sake of illustration, let us first consider the case of one single se\-parable potential supporting the ground state of atomic hydrogen. In this case, the choice of the separable potential is unique. We have:
\begin{equation}
V(\vec{p},\vec{p}')=v_{1s}(\vec{p})v_{1s}^*(\vec{p}'),
\end{equation}
where,
\begin{equation}
v_{1s}(\vec{p})=\frac{4\sqrt{\pi}}{p^2+1}.
\end{equation}
In momentum space, the action of this separable potential on the state vector $|\mathbf{\Phi}\rangle$ is:
\begin{equation}
\hat{V}|\mathbf{\Phi}\rangle=-\frac{16\pi}{p^2+1}\int\frac{\mathrm{d}\vec{p}'}{(2\pi)^3}\frac{\mathbf{\Phi}(\vec{p}')}{p^{'2}+1}.
\end{equation}
By inverse Fourier transform, we obtain the action of the same separable potential on the state vector $|\mathbf{\Psi}\rangle$ in the configuration space:
\begin{equation}
\hat{V}|\mathbf{\Psi}\rangle=-\frac{1}{\pi}\frac{e^{-r}}{r}\int\mathrm{d}\vec{r}\mathbf{\Psi}(\mathbf{r}')\frac{e^{-r'}}{r'}.
\end{equation}
We therefore see that, in this case, the separable potential reduces to a nonlocal Yukawa potential of short range in the configuration space. It is now interesting to generate the separable potentials which support the 1s and 2s states. In that case, the choice is not unique. We have:
\begin{equation}
V(\vec{p},\vec{p}')=v_{1s}(\vec{p})v_{1s}^*(\vec{p}')+v_{2s}(\vec{p})v_{2s}^*(\vec{p}').
\end{equation}
From Eq. (13), we have:
\begin{multline}
v_{1s}(\vec{p})=\alpha_{11}\left(\varepsilon_{1s}-\frac{1}{2}p^2\right)\varphi_{1s}(\vec{p})\\
+\;\alpha_{12}\left(\varepsilon_{2s}-\frac{1}{2}p^2\right)\varphi_{2s}(\vec{p}),
\end{multline}
\begin{multline}
v_{2s}(\vec{p})=\alpha_{21}\left(\varepsilon_{1s}-\frac{1}{2}p^2\right)\varphi_{1s}(\vec{p})\\
+\;\alpha_{22}\left(\varepsilon_{2s}-\frac{1}{2}p^2\right)\varphi_{2s}(\vec{p}).
\end{multline}
The $\alpha_{ij}$ coefficients which are the elements of the matrix $\mathbf{A}^{-1}$, may be obtained by solving Eq. (14) after imposing the prescription described above. It is important to stress the following point. If the separable potentials support $N$ bound states of atomic hydrogen, the wave function associated to each of these bound states is exact. For the continuum states however, this is not the case since in the configuration space, the separable potentials have a short range. These wave functions associated to the continuum are calculated in Section 2.5.\\

We introduce here a second method to generate the separable potentials which is based on the following expansion of the Coulomb potential kernel:
\begin{equation}
\label{eq_pot_decomp}
V(\vec p-\vec u)=\sum_{\ell=0}^\infty\sum_{n=0}^\infty N_{nl}v_{n\ell }^*(p)v_{n\ell }(u)\sum_{m=-\ell}^\ell Y_{\ell m}^*(\Omega_p)Y_{\ell m}(\Omega_u).
\end{equation}
$v_{nl}(p)$ is expressed in term of a Gegenbauer polynomial: 
\begin{equation}
v_{nl}(p)=\frac{1}{p}\left(\frac{2qp}{q^2+p^2}\right)^{l+1}C_n^{l+1}\left[\frac{q^2-p^2}{q^2+p^2}\right],
\end{equation}
and $Y_{\ell m}(\Omega)$ is a spherical harmonic. The coefficient $N_{nl}$ is given by:
\begin{equation}
N_{nl}=\pi[\Gamma(l+1)]^2\frac{2^{2l+2}(2l+1)}{\Gamma(n+2l+2)}n!.
\end{equation}
Expansion (23) follows from Eq. (2) in \cite{Cohl} and Eq. (11) in \cite{Ossicini}.
In practice, of course, expansion (23) is truncated. It is interesting to note that $v_{nl}$ is nothing else than the usual Coulomb sturmian function in momentum space. This function contains the free parameter $q$. For a given value of this parameter, the expansion (23) is unique. If we perform the inverse Fourier transform of  function $v_{nl}(p)$, we obtain the Coulomb sturmian function in the confi\-guration space up to a normalization factor. In this way, it becomes clear that the parameter $q$ determines the range of the potential. Tuning $q$ for each $\ell$ in Eq. (23) allows one to adjust the potential range.

\subsection{The Volterra integral equation}
For the sake of completeness, let us now analyze the Volterra equation (10) in the simplest case of one separable potential supporting the 1s state. By writing $x=(t-\xi)/2$ and $y=b(t)-b(\xi)$, the kernel $K(x,y)$ may be written as:
\begin{equation}
K(x,y)=\varepsilon_{1s}\varepsilon_{1s} J(x,y)-\varepsilon_{1s}H(x,y)+\frac14L(x,y)
\end{equation}
where
\begin{equation}
J(x,y)=\int\frac{\mathrm{d}\vec{p}}{(2\pi)^3}\varphi_{1s}(\vec p)\varphi_{1s}(\vec p)e^{-\mathrm{i}xp^2+\mathrm{i}y(\vec e\cdot\vec p)},
\end{equation}
\begin{equation}
H(x,y)=\int\frac{p^2\mathrm{d}\vec{p}}{(2\pi)^3}\varphi_{1s}(\vec p)\varphi_{1s}(\vec p)e^{-\mathrm{i}xp^2+\mathrm{i}y(\vec e\cdot\vec p)},
\end{equation}
\begin{equation}
L(x,y)=\int\frac{p^4\mathrm{d}\vec{p}}{(2\pi)^3}\varphi_{1s}(\vec p)\varphi_{1s}(\vec p)e^{-\mathrm{i}xp^2+\mathrm{i}y(\vec e\cdot\vec p)}.
\end{equation}
$F_0(t)$ can be written in terms of the previous functions:
\begin{equation}
F_{0}(t)=\varepsilon_{1s}J(x,y)-\frac12H(x,y),
\end{equation}
where this time, $x=t/2$ and $y=b(t)$. There are va\-rious ways to calculate numerically the functions $J(x,y)$, $H(x,y)$ and $L(x,y)$. One way to proceed is to use recursion formulae as described in \cite{Tetchou}. If the separable potentials support more than one bound states, the kernel becomes a matrix and $F_0(t)$ becomes a vector. In that case, the elements of the matrix kernel can be expressed in terms of functions similar to $J(x,y)$, $H(x,y)$ and $L(x,y)$ but involving the wave functions corresponding to the bound states included in the model.This remark applies also to the components of vector $\mathbf{F}_0$. 

\subsection{Calculating observables}
The probability amplitude for the system to stay, at the end of the interaction,  in one of the bound states of atomic hydrogen taken into account in the model can be calculated by projecting the final wave packet on the corresponding wave function:
\begin{equation}
\langle \varphi_{nl}|\mathrm{\Phi}(T) \rangle=\sum_{m=-\ell}^{\ell}\int\frac{\mathrm{d}\vec{p}}{(2\pi)^3}\varphi_{nl}(p)Y^*_{lm}(\vec p)\mathrm{\Phi}(\vec p,T)
\end{equation}
The total ionisation probabilty can be defined in two ways: (i) as one minus the amount of  population staying in all the bound states after the interaction, or (ii) via the integration of the energy spectrum. If, for instance, the model includes $N_s$ s-states, $N_p$ p-states and $N_d$ d-states, the ionisation probability is given in the first case by:
\begin{multline}
\label{eq_iy_pw}
P_{\mathrm{ion}}(T)=1-\sum_{n}^{N_s}|\langle \varphi_{ns}|\mathrm{\Phi}(T) \rangle|^2\\
-\sum_{n}^{N_p}\sum_{m=-1}^1|\langle \varphi_{np,m}|\mathrm{\Phi}(T) \rangle|^2-\sum_{n}^{N_d}\sum_{m=-2}^2|\langle \varphi_{nd,m}|\mathrm{\Phi}(T) \rangle|^2.
\end{multline}
This last expression is only valid if the separable potentials have been generated by imposing that they support these bound states. In the case where the separable potentials are generated by using Eq. (23), we do not know, {\it a priori} how many bound states are well described in the model. The only way of calculating the ionisation probability in that case, is to integrate the energy spectrum:
\begin{equation}
\label{eq_iy_cw}
P_{\mathrm{ion}}(T)=\int_0^\infty \frac{\mathrm{d} P(T)}{\mathrm{d} E_k}\mathrm{d}E_k, 
\end{equation}
where, 
\begin{equation}
 \frac{\mathrm{d} P(T)}{\mathrm{d} E_k}= \int|C(\vec k,T)|^2\mathrm{d}\Omega.
\end{equation}
$C(\vec k,T)$ represents the amplitude of probability for the electron to be in the continuum at the end of the interaction ($t=T$) with a velocity $\vec{k}$. It is defined as:\begin{equation}
C(\vec k, T)=\langle \varphi^-_{\vec{k}}|\mathrm{\Phi}(T)\rangle=\int\frac{\mathrm{d}\vec{p}}{(2\pi)^3}\varphi^{-*}_{\vec{k}}(\vec p)\mathrm{\Phi}(\vec p,T).
\end{equation}
The continuum wave functions $\varphi^-_{\vec{k}}(\vec{p})$ which behave asymptotically as an incoming spherical wave are a solution of the following eigenvalue equation:
\begin{equation}
\frac{p^2}{2}\varphi^-_{\vec{k}}(\vec{p})-\sum^N_{n=1}v_n(\vec{p})\int\frac{\mathrm{d}\vec{u}}{(2\pi)^3}v_n^*(\vec{u})\varphi_{\vec{k}}^-(\vec{u})=\frac{k^2}{2}\varphi_{\vec{k}}^-(\vec{p}).
\end{equation}
It is expressed as follows:
\begin{multline}
\label{eq_exact_cont}
\varphi^-_{\vec{k}}(\vec p)=(2\pi)^{3/2}\delta(\vec p-\vec k)-\\
-\frac{2}{(k-\mathrm{i}\varepsilon)^2-p^2}\int\frac{\mathrm{d}\vec{u}}{(2\pi)^3}V(\vec p,\vec u)\varphi^-_{\vec{k}}(\vec u).
\end{multline}
Note that in the rhs of Eq. (37), the first term is a plane wave and the second term is a correction resulting from the short range separable potential $V(\vec p,\vec u)$. \\

\subsection{Gauge invariance}
Contrary to what is claimed in \cite{Tetchou}, the formulation of the model is not gauge invariant. This is true irrespective of the method used to generate the separable potentials unless, of course, all the terms of expansion (23) are taken into account in the model.\\

The problem of the gauge invariance in the context of nonlocal potentials is rather subtle. Following Korolev \cite{Korolev}, a nonlocal potential in the configuration space can be written as a local potential depending on $\vec{r}$ and $\vec{p}$. To show it, we write:
\begin{eqnarray}
\Psi(\vec{r}')&=&\left(\sum_n \frac{1}{n!}(\vec{r}'-\vec{r})^n\cdot\vec{\nabla}_{\vec{r}}^n\right)\Psi(\vec{r})\nonumber\\
&=&\left(\sum_n \frac{1}{n!}(\vec{r}'-\vec{r})^n\cdot(i\vec{p})^n\right)\Psi(\vec{r})\nonumber\\
&=& \hat{S}\left((\vec{r}'-\vec{r})\cdot\vec{p}\right)\Psi(\vec{r}),
\end{eqnarray}
where the momentum $\vec{p}\equiv -i\vec{\nabla}_{\vec{r}}$.
At this stage, it is important to note that the operator $ \hat{S}\left((\vec{r}'-\vec{r})\cdot\vec{p}\right)$ is not ri\-gorously equal to the translation operator $\exp{[i(\vec{r}'-\vec{r})\cdot\vec{p}]}$ since this ope\-rator acting on a function translates it by a quantity which is $\vec{r}$-dependent \cite{remark1}. Now, if $V$ is a nonlocal potential, we can write in the configuration space:
\newpage
\begin{multline}
\int\langle\vec{r}|V|\vec{r}'\rangle\Psi(\vec{r}')d\vec{r}'\nonumber\\
=\left\{\int\langle\vec{r}|V|\vec{r}'\rangle \hat{S}\left((\vec{r}'-\vec{r})\cdot\vec{p}\right)d\vec{r}'\right\}\Psi(\vec{r})\nonumber
\end{multline}
\begin{equation}
\equiv V(\vec{r},\vec{p})\Psi(\vec{r}).\;\;\;\;\;\;\;\;\;\;\;\;\;\;\;\;\;
\end{equation} 
where the local momentum-dependent potential $V(\vec{r},\vec{p})$ is given by the expression in brackets. \\

If we work then in  configuration space, we see that our Hamiltonian will contain two terms depending on the momentum: the kinetic ener\-gy operator and the potential operator. It means that if we consider the interaction of the electron with the electromagnetic field, it is necessary to perform the minimal substitution $\vec{p}\rightarrow\vec{p}+\vec{A}/c$ in both terms to obtain a fully gauge-invariant formulation. However, moving to the momentum space leads, in this case, to a much more complicated equation than Eq. (2), the solution of which is no longer given by Eq. (9). In addition, the fact that the atomic potential is now depending on the external fields is often considered as non physical \cite{remark2}. This point has been stressed recently by Rensink and Antonsen \cite{Rensink}. They claim that a model involving a nonlocal potential is intrinsically gauge dependent and show that in the case of the linear field response, the length gauge is preferable for nonlocal atomic binding potentials. In the present case however, we are interested in the nonlinear field response. We calculated the electron energy spectrum for frequencies slightly higher than the ionisation potential where the number of bound states playing a significant role is small. We performed the calculations (i) by using a fully gauge inva\-riant formulation, (ii) by using our approach developed in Section (2) in the velocity gauge and (iii) by solving numerically the TDSE with the full Coulomb potential. It turns out that our approach in the velocity gauge gives results that are in much better agreement with the TDSE results obtained with the full Coulomb potential than the results obtained by using the gauge invariant formulation of our model.\\

The problem of the existence of a preferable gauge has been studied for a long time in the context of the SFA \cite{Bauer}. Various SFA calculations have been performed in the length and velocity gauge. The comparison of these results with those obtained with TDSE has shown that the "best" gauge depends strongly on what we are calcula\-ting and on the field parameters. In other words, it doesn't make a lot of sense to speak about a preferable gauge in that case. In order to solve this problem in the case of SFA, we developed recently a new approach. We showed that all SFA schemes may be grouped into a set of different families. Within a family, it is possible to define a length and a velocity gauge scheme that give identical results. However, irrespective of the gauge, schemes belonging to different families give different results. This shows clearly the non-existence of a privileged gauge. In order to define these families, we consider an ansatz to describe the wave packet. It consists in defining the electron wave packet as the sum of the initial state wave function times a phase factor and a function which is the solution of an inhomogeneous TDSE. It is the phase factor that characterizes a given family. In his attempts to derive a gauge invariant formulation of the SFA, Faisal \cite{Faisal} arrived at similar results.\\

It turns out that in the case of our separable potential model,  it is also possible to define two families of formulations of the TDSE. In the theory involving a local potential, the wave functions in the length and velocity gauge are connected with the well known G\"oppert-Mayer phase factor in the configuration space. In the momentum space, this connection takes the following form:
\begin{equation}
\Psi_V(\vec{p},t)=\Psi_L(\vec{p}-b'(t)\vec{e},t).
\end{equation}
Now, instead of writing the Coulomb potential like in Eq. (3), we can also write:
\begin{multline}
V(\vec{p}-\vec{u})=V(\;(\vec{p}-b'(t)\vec{e})-(\vec{u}-b'(t)\vec{e})\;)\\
\approx\sum_{n=1}^Nv_n(\vec{p}-b'(t)\vec{e})v^*_n(\vec{u}-b'(t)\vec{e}).
\end{multline}
On the other hand, Eq. (2) is written in the velocity gauge, which means that $\Psi(\vec{p},t)\equiv\Psi_V(\vec{p},t)$. Inserting now expression (41) into (2), we obtain after simple algebraic manipulations the following equation for the wave function in the length gauge:
\begin{multline}
\left[i\frac{\partial}{\partial t}-\frac{p^2}{2}+ib''(\vec{e}\cdot\vec{\bigtriangledown_{\vec{p}}})\right]\Psi_L(\vec{p},t)\\
+\int\frac{\mathrm{d}\vec{u}}{(2\pi)^3}V(\vec{p},\vec{u})\Psi_L(\vec{u},t)=0.
\end{multline}
So, the separable potentials (3) and (41) form one gauge invariant family. Another family is formed if we use the separable potential (3) in Eq. (42) and the potential (41) in the Eq. (2) replacing in the potential $b'(t) \rightarrow -b'(t)$ in agreement with the inverse G\"oppert-Mayer gauge transformation. Of course, within a family, results for any physical observable are identical. Nevertheless, the "global" gauge invariance is a problem for separable potentials.

\subsection{The strong field approximation}
The SFA theory has been essentially used to study the interaction of one-active electron systems with a strong laser pulse. In the low frequency regime ({\it i.e.} when the ionisation potential is much bigger than the photon energy), this interaction leads to two important competing processes namely the ionisation of the atom and the generation by the atom of high order harmonics of the driving field. Within the SFA, the underlying electron dynamics takes place in three steps. The first step is a one-pseudophoton transition from the bare atomic ground state to a dressed continuum which is a Volkov state. This first step is usually interpreted in terms of electron tunneling  for reasonably strong fields \cite{Lewenstein1,Lewenstein2}. Once the electron is emitted, it undergoes a quiver motion driven by the oscillating field. During this motion, the Coulomb binding potential is neglected except for electrons returning to the nucleus when they either recombine into the ground state thereby  leading to the emission of high order harmonics, or when they undergo an elastic scattering by the Coulomb potential of the residual ion. Let us note that to describe within the SFA, the recombination of electrons into the ground state requires the calculation of the full time-dependent atomic dipole. \\

The electrons that do not return to the nucleus are called direct electrons. Their dynamics in the oscillating field is described by what we call the first order term of the SFA expansion which is nothing else than the usual SFA approach. Electrons that undergo a quiver motion are indirect electrons. They experience one or more elastic re-scatterings by the residual ion that are described by the high-order terms of the SFA expansion.  At this stage, it is important to stress that within the SFA, no excited states of the atomic binding potential are  taken into account.\\

We now consider our model in which we assume that our separable potential supports only the ground atomic state. In this case, we know that this potential reduces to a separable Yukawa potential of short range in the configuration space (see Eq. (19) for the case of atomic hydrogen). In these conditions, we expect our model to describe the same type of electron dynamics as in the SFA. In fact, the first order term of the SFA expansion may be obtained as a limiting case of  our model. \\

We consider the case of atomic hydrogen and start from Eq. (2) in which we replace the kernel of the nonlocal Coulomb potential by $V(\vec{p},\vec{u})$ given by Eq. (16):
\begin{equation}
\left [\mathrm{i}\frac{\partial}{\partial t}-\frac{p^2}{2}+b'(t)(\vec e\cdot\vec p)-\zeta'(t)\right]\Psi(\vec p,t) + \frac{4\sqrt{\pi}}{p^2+1}F(t)=0,
\end{equation}
where we used Eq. (4) and (5). $F(t)$,  given by Eq. (8) where we omit the subscript $n$, since the separable potential contains only one term, becomes:
\begin{equation}
F(t)=\frac{4}{\sqrt{\pi}}\int \frac{\mathrm{d}\vec{u}}{(2\pi)^3}\frac{\Psi(\vec{u},t)}{(u^2+1)}.
\end{equation}
Note that in order to establish the link with the SFA, we have to keep the $A^2(t)$ term in Eq. (43). The solution of Eq. (43) can be written formally as:
\begin{multline}
\Psi(\vec{p},t)=e^{-\mathrm{i}\frac{p^2}{2}t+\mathrm{i}b(t)p_z-\frac{\mathrm{i}}{2}\zeta(t)}\;[\;\;\Psi(\vec{p},0)\;+\\
+\frac{4\mathrm{i}\sqrt{\pi}}{(p^2+1)}\int^t_0F(\xi)e^{\mathrm{i}\frac{p^2}{2}\xi-\mathrm{i}b(\xi)p_z+\frac{\mathrm{i}}{2}\zeta(\xi)}\mathrm{d}\xi\;\;] ,
\end{multline}
where,
\begin{equation}
\Psi(\vec{p},0)=\psi_{1s}(p)=\frac{8\sqrt{\pi}}{(p^2+1)^2},
\end{equation}
is the atomic hydrogen ground state wave function in the momentum space. If we replace $\Psi(\vec{u},t)$ in Eq. (44) by $\exp(-i\varepsilon_{1s}t)\varphi_{1s}(\vec{u})$, expression (45) for the total wave packet reduces exactly to the corresponding first order term of the SFA expansion in the velocity gauge  and in the first family (see \cite{Galstyan}):
\begin{multline}
\Psi_{V-\mathrm{SFA}}^{(1)}(\vec{p},t)=e^{-\mathrm{i}\frac{p^2}{2}t+\mathrm{i}b(t)p_z-\frac{\mathrm{i}}{2}\zeta(t)}\;\psi_{1s}(p)\;\;[1+\\
+\mathrm{i}(\frac{p^2}{2}-\varepsilon_{1s})\int^t_0e^{\mathrm{i}(\frac{p^2}{2}-\varepsilon_{1s})\xi-\mathrm{i}b(\xi)p_z+\frac{\mathrm{i}}{2}\zeta(\xi)}\mathrm{d}\xi\;\;] .
\end{multline}

\section{Results and discussion}
We use our model to study the interaction of atomic hydrogen with a sine square linearly polarized radiation pulse at a wave length of 800 nm. We have two objectives: first,  to analyze the role of the excited states in the ionisation process and second, to assess the limits of validity of the SFA. Here, we mainly study the modulus square of the total wave packet in momentum space as a function of both, the time and the canonical momentum $p_z$ along the polarization axis (which coincides with the z-axis) while keeping constant $p_n$, the canonical momentum in the plane perpendicular to the polarization axis. For the sake of comparison, we start with results shown in Fig. 1 and obtained by solving the full TDSE in which the Coulomb potential is fully taken into account. We consider a 10 cycle pulse of peak intensity $I=4\times 10^{14}$ W/cm$^2$ and analyze the wave packet in a direction parallel to the polarization axis ($p_n=0$). The electron probability density in momentum space shown in Fig.1 exhibit 
\begin{figure}
\includegraphics[width=0.5\textwidth]{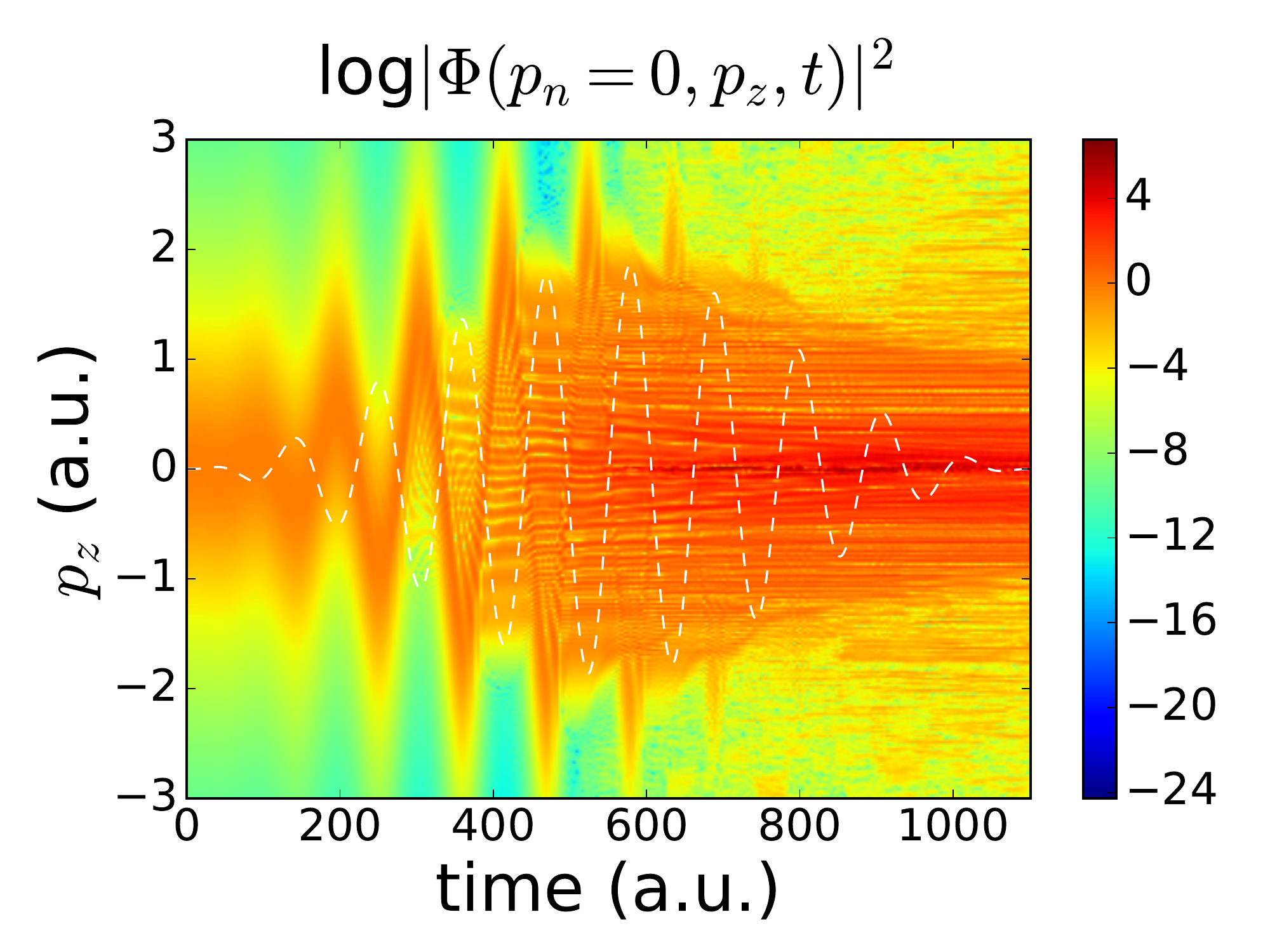}
\caption{Logarithm of the electron probability density in momentum space as a function of both the time and the canonical momentum along the polarization axis $p_z$. The canonical momentum in the plane perpendicular to the polarization axis $p_n=0$. The electron probability density is obtained by solving the TDSE describing the interaction of atomic hydrogen with a linearly polarized 10 cycle sine square pulse of frequency $\omega=0.057$ a.u. and peak intensity $I=4\times10^{14}$ W/cm$^2$ . The white dashed line represents the vector potential $A(t)$ which oscillates in phase opposition to the bound state population.}
\label{fig_tdse_pn0}      
\end{figure}
oscillations and horizontal stripes. Since, for bound states,  the electron averaged velocity along the z-axis $\langle v_z\rangle_{\mathrm{bound}}\approx 0$, we have $\langle p_z\rangle_{\mathrm{bound}}\approx -A(t)$. In other words, the bound state component of the wave packet in the momentum space oscillates in phase opposition to the vector potential (dashed line in Fig.1). According to Ehrenfest's theorem  (see Eq. (1)), the canonical momentum becomes a constant of motion that reduces to the electron drift velocity, in atomic units, once it is ionized. Therefore, the stripes represent the continuum state components of the wave packet in the momentum space. These stripes originate from the wave packet bound state components at time $t=t_{\mathrm{ionisation}}$ such that:
\begin{equation}
p_z(t_{\mathrm{ionisation}})=-A(t_{\mathrm{ionisation}})=v_z^{\mathrm{final}}.
\end{equation}
When they start, the stripes are rather thick. They become thinner after each new optical cycle. This feature is a signature of the inter-cycle interferences \cite{Arbo}. Once it is ionized, a wave packet can interfere with another wave packet emitted one optical cycle before and propagating in the same direction. In other words, the stripes are interference fringes. This feature is  much more pronounced in the case of our model when only a small number of bound states are included. Finally, let us mention that some of the stripes characterized by a small value of $|p_z|$ are not horizontal but get closer to the line $p_z=0$. This effect results from the Coulomb focusing of the slow electron wave packets. It is important to stress that in the present case where the full Coulomb potential is taken into account, the problem is gauge invariant and yet, there is a clear separation between bound and continuum components of the wave packet in the momentum space. Although this representation of the electron probability density is not directly observable, it provides information on the ionisation time and on the ionisation mechanism. In this case, we see in Fig. 1 that the strongest stripes are characterized by small values of $|p_z|$ around $A(t)=0$, {\it i.e.} at the maximum of the electric field as expected in the present regime of long wave lengths.\\

\begin{figure}[!h]
\includegraphics[width=0.5\textwidth]{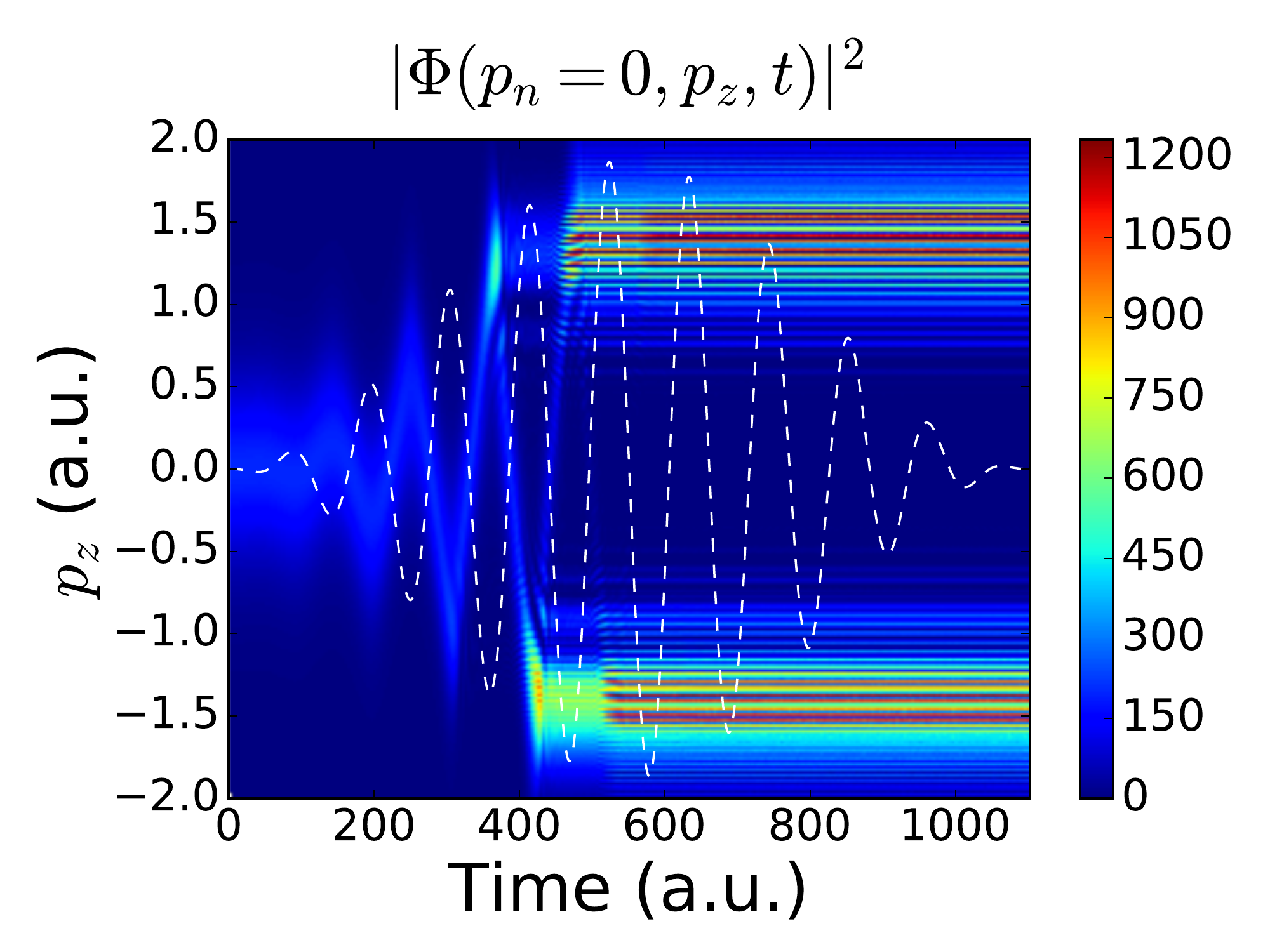}
\caption{Electron probability density in momentum space as a function of both the time and the canonical momentum along the polarization axis $p_z$. The canonical momentum in the plane perpendicular to the polarization axis $p_n=0$. This electron probability density is obtained within our model where only the 1s state is taken into account and for the same case as in Fig. 1. As in Fig. 1, the dashed line represents the vector potential $A(t)$.}
\label{fig_1s_pn0}      
\end{figure}
In Fig. 2, we show the electron probability density obtained, for the same case as before, by means of our model supporting only the 1s state. We clearly see the effect of the inter-cycle interference between 400 and 600 atomic units of time around $p_z=-1.5$ a.u.. The most striking result however, is the fact that the electrons are emitted when the vector potential is at its maximum {\it i.e} when the electric field is close to zero. It means that, quite unexpectedly, tunneling is not the dominant ionisation mechanism thereby preventing the emission of slow electrons in the direction of the polarization axis. The canonical momentum of the emitted electrons $p_z\approx -A_0$ where $A_0$ is the amplitude of the vector potential. Within the dipole approximation, the only way for the electrons to acquire such momentum is to make a (laser assisted) inelastic collision with the residual ion. In addition, the kinetic energy $E_K$ of the emitted  electrons is given by:
\begin{equation}
E_K=\frac{p_z^2}{2}=\frac{A_0^2}{2}=2U_p,
\end{equation}
where we used Eq. (6). $U_p$ is the ponderomotive potential defined in terms of the electric field amplitude $E$ by $E^2/(4\omega^2)$. Note that this result seems compatible with the "simple man's model" based on the classical theory which describes the electrons once they are emitted and in which the electron momentum at the end of the pulse is determined by the value of the vector potential at the time of ionisation provided that these electrons are direct {\it i.e.} do not come back to the residual ion. It is important to mention also that for the peak intensity considered here ($I=4\times 10^{14}$ Watt/cm$^2$), over-the-barrier ionisation instead of tunneling could be expected to be the dominant ionisation mechanism. We have checked that all the features we observe in this case are also present at much lower peak intensity. Therefore, when only the ground state is taken into account in our model or, in other words, when the atomic binding potential has a short range like for a negative ion, it becomes clear that the dominant process of emission of electrons results from a laser assisted inelastic collision of these electrons with the residual ion and not from tunneling. The first order SFA  predicts under the same conditions that tunneling is the dominant ionisation process leading preferentially to the emission of very slow electrons. Since the first order SFA only describes the emission of direct electrons and not the re-scattered electrons, the differences with the predictions of the model could indicate that the dominant contribution to the ionisation process comes from the indirect electrons.\\

\begin{figure}
\includegraphics[width=0.5\textwidth]{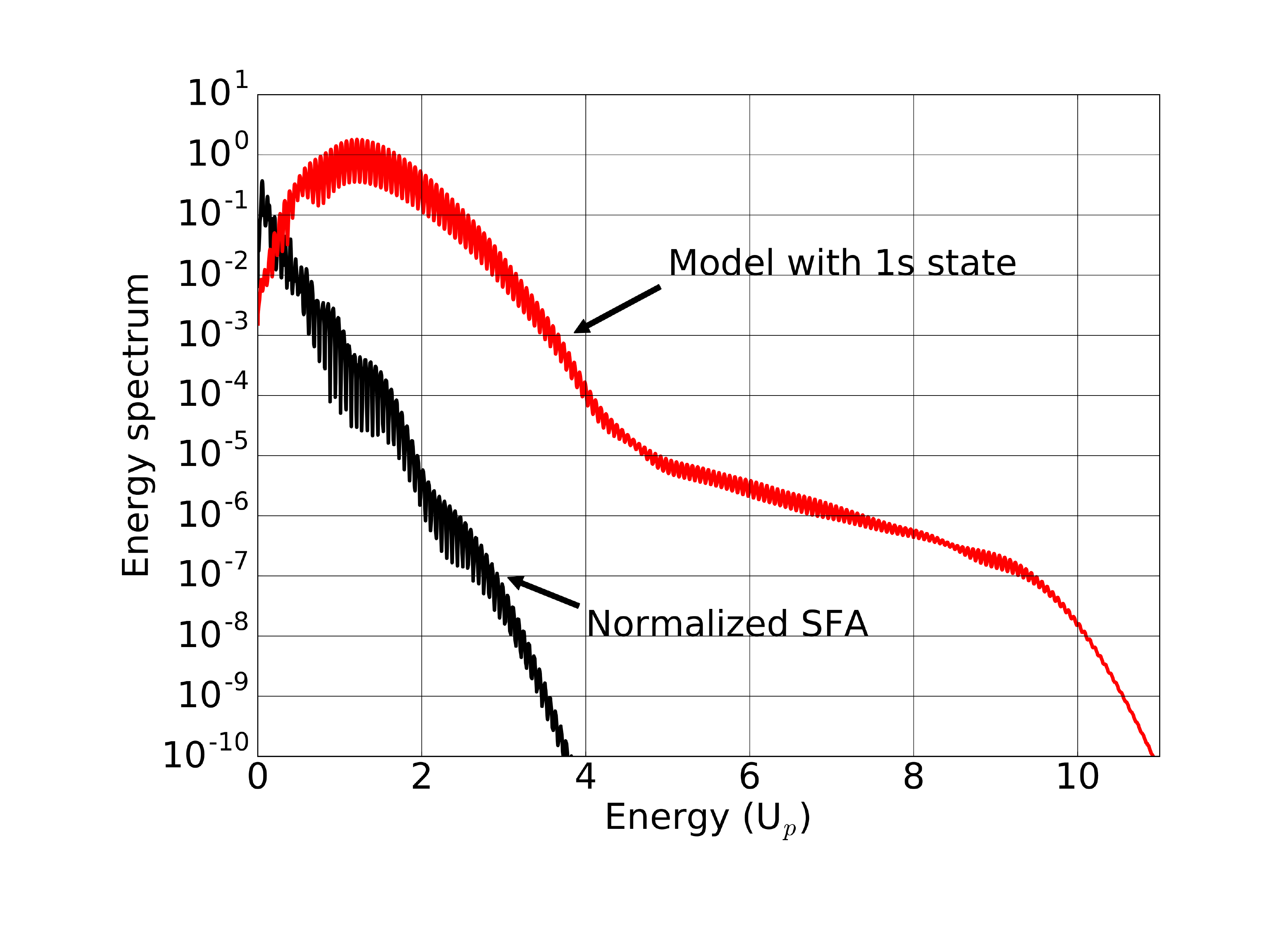}
\caption{Electron energy spectrum (integrated over the angles) for the same case as in Fig. 1. The red curve is obtained by using our model which only takes into account the ground state of atomic hydrogen. The black curve is the SFA result. }
\label{fig_geg_mmap_n1l1}      
\end{figure}

In Fig. 3, we show the results for the electron energy spectrum integrated over the angles and for the same case as in Fig. 1. The spectrum obtained by using our model in which only the 1s state is taken into account, exhibits a maximum around $2U_p$. This maximum is followed by a "plateau" of much lower amplitude and that extends until about $10U_p$. This result is compatible with the "simple man's model" . However, it differs from the spectrum obtained within the regular SFA which exhibits a maximum at very small electron energy followed by a rapid decrease.\\

In Fig. 4, we consider the logarithm of the electron probability density, obtained with our model that takes only the 1s state into account,  for the same laser parameters as before but the component $p_n$ of the canonical momentum in the plane perpendicular to the polarization axis, is fixed and equal to 1.2 in atomic units. 
\begin{figure}
  \includegraphics[width=0.5\textwidth]{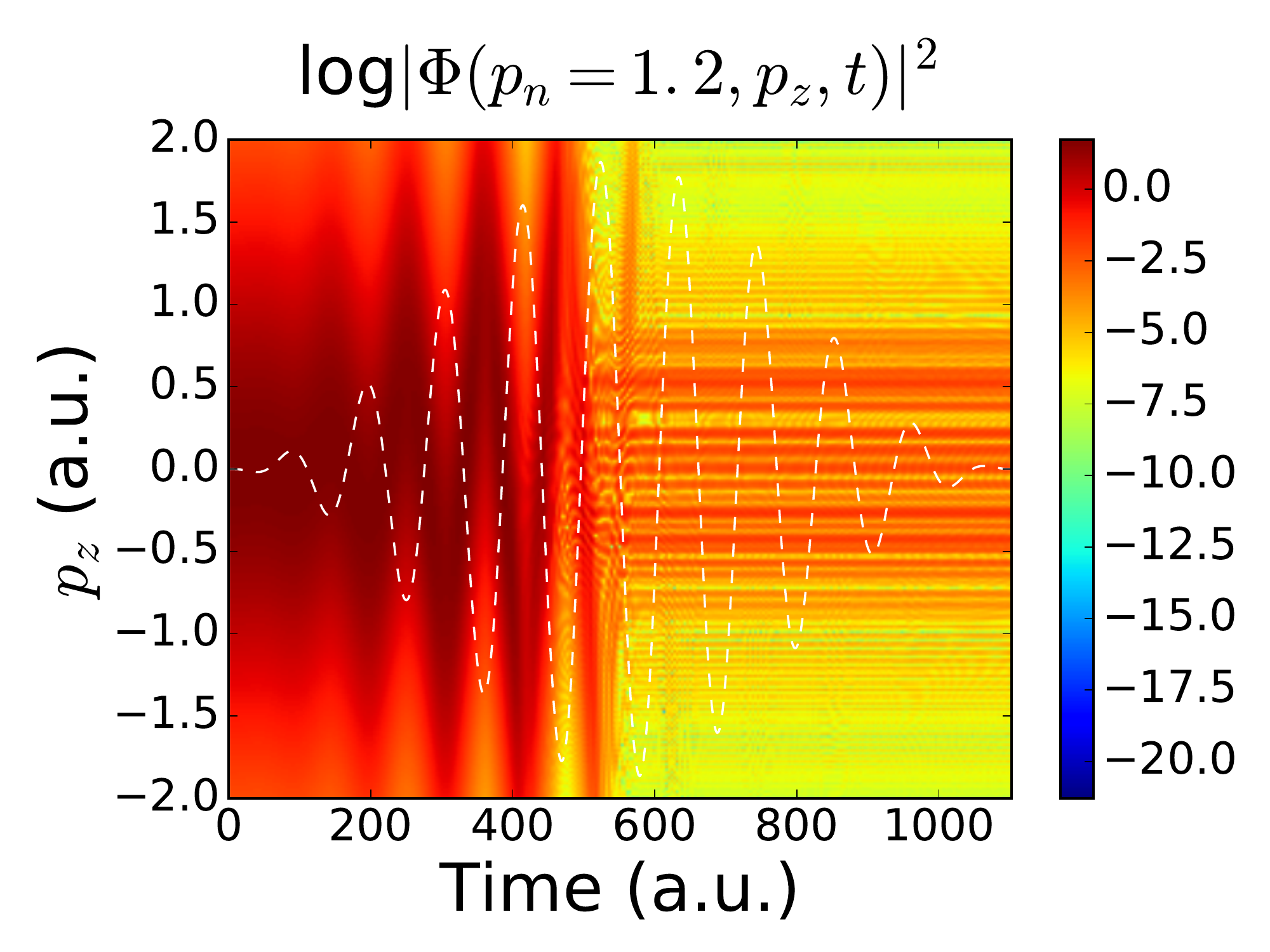}
\caption{Logarithm of the electron probability density for the same laser parameters as in Fig. 1 but, the components of the canonical momentum $p_n$ in the plane 
perpendicular to the polarization axis is fixed and equal to 1.2 in atomic units. This result is obtained with our model in which only the 1s state is taken into account.}
\label{fig_1s_pn12}       
\end{figure}
In that case, the magnitude of the probability density is much lower. We clearly see that beyond 400 a.u.  of time after the beginning of the pulse, no population is left in the ground state. It is interesting to see that most of the stripes cor\-respond to very small values of $|p_z|$. 
\begin{figure}
\includegraphics[width=0.5\textwidth]{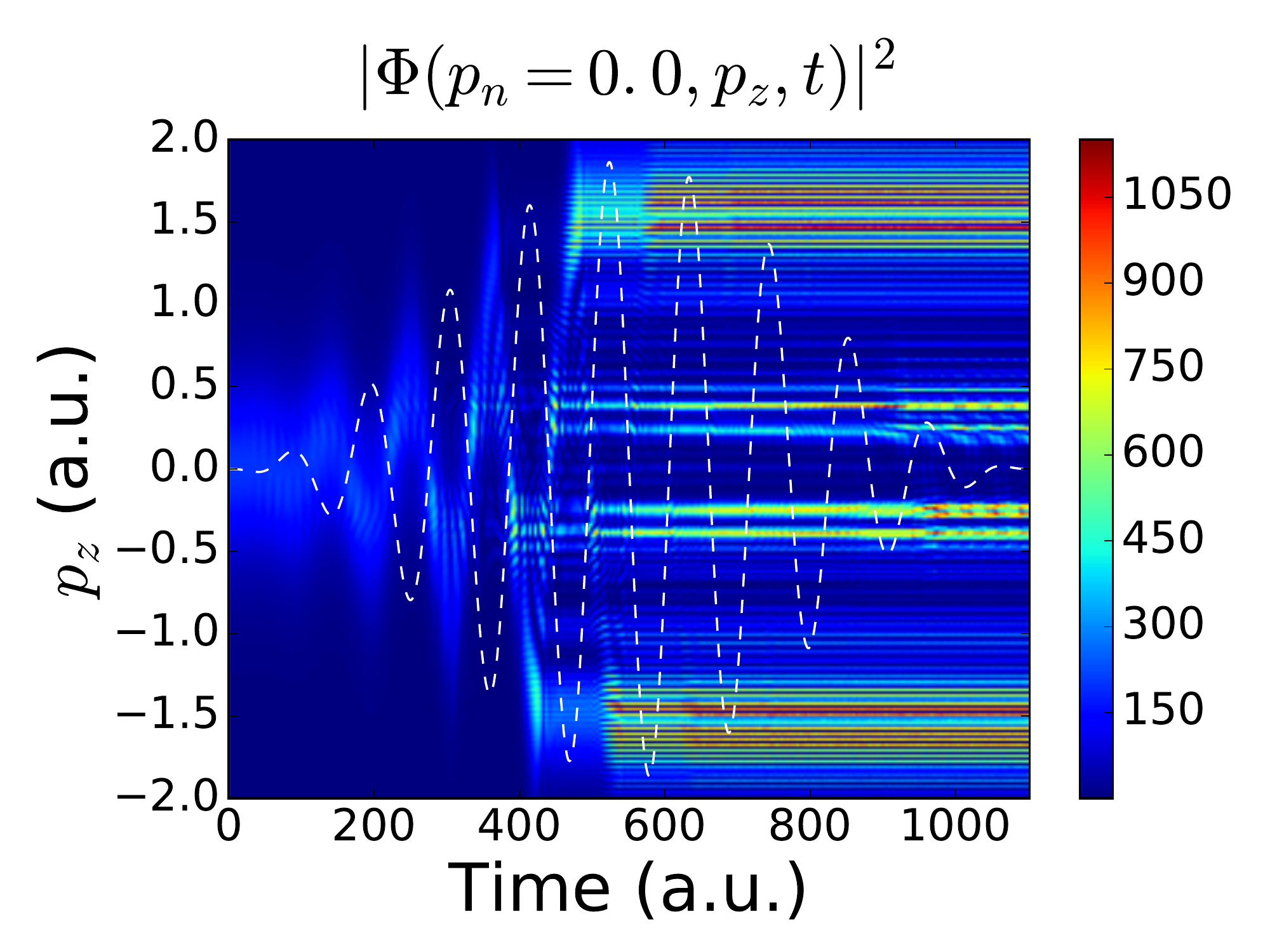}
\caption{Electron probability density in momentum space as a function of both the time and the canonical momentum along the polarization axis $p_z$. The canonical momentum in the plane perpendicular to the polarization axis $p_n=0$. This electron probability density is obtained within our model where the 1s, 2p and 3d states are taken into account and for the same case as in Fig. 1. As in Fig. 1, the dashed line represents the vector potential $A(t)$.}
\label{fig_1s2p3d_pn0}      
\end{figure}
These stripes are associated with electrons that are emitted when the field is ma\-ximum and in a direction perpendicular to the polarization axis. For the present peak intensity ($4\times 10^{14}$ W/cm$^2$), the underlying ionisation mechanism could be over-the-barrier ionisation or again a laser-assisted inelastic collision with the residual ion.\\

Let us now examine the behaviour of the electron probability density  when, in our model, we take into account several bound states of atomic hydrogen, namely the 1s, 2p and 3d states and therefore extending the range of our potential. Note that in this specific case, the choice of the corresponding separable potentials is unique. The laser parameters are the same as before and we consider the emission of electrons in a direction parallel to the polarization axis ($p_n=0$). Beside the stripes associated to electrons emitted at the maximum of the vector potential, new stripes appear, corresponding to low energy electrons emitted when the electric field is near its maximum. We have checked that these stripes are also present for lower peak intensities. It is important to note that the ionisation yield decreases systematically as soon as other states than the 1s state are included. This may be understood in terms of a polarization effect. For a model atom in which only the 1s state is included, only the continuum state contributes to the polarization of the electron cloud in the presence of the field, thereby leading to a rapid ionisation. When, in our model, we include excited bound states and in particular, p-states, they will contribute significantly to the polarization of the electronic cloud without necessarily leading to ionisation unless the field becomes very strong.

\section{Conclusions and perspectives}
In this contribution, we have studied in depth and extended an existing model describing the interaction of a one-active electron system with a strong laser field. The model is based on the expansion of the kernel of the binding potential in momentum space in a finite sum of $N$ separable potentials. This approach requires the solution of a system of $N$ one-dimensional Volterra integral equations instead of a four-dimensional TDSE.  Here, we focused on three important aspects namely the different ways of generating the separable potentials, the  gauge inva\-riance of the model and the link existing between this model and the usual SFA.\\

The separable potentials may be generated by imposing that each of them supports one bound state of the binding potential. In the case where the binding potential is a pure Coulomb potential, the corresponding kernel can also be expanded in products of separable potentials expressed in terms of Gegenbauer polynomials. These separable potentials depend on a free parameter. By moving to the configuration space, we have shown that this parameter determines the range of each of these separable potentials.\\

 In principle, the TDSE associated with our model can be formulated in a fully gauge invariant way provided that the separable potentials depend on the external fields. However, the fact that the separable potentials are field dependent has been considered as unphysical. As a matter of fact, our original formulation which is gauge dependent, turned out to give results for the electron energy spectra in very good agreement with the TDSE results obtained by treating the binding potential exactly in cases where the number of states playing an important role is relatively small. This agreement is not so good if we use the fully gauge invariant formulation instead. The question of a preferable gauge in this context has also been discussed in a way analogous to what we did recently in the case of the SFA.\\ 

In the case where the model takes into account only one separable potential which supports the ground state of the atom, we have shown that the usual SFA can be obtained from the model in a well defined limit.\\

We have  applied our model to the calculation of the electron probability density in momentum space at a wave length of 800 nm. The study of such probability density as a function of both the time and the canonical momentum along or perpendicular to the polarization axis provides valuable information on the ionisation times and on the ionisation mechanism. In particular, we have shown that by contrast to the predictions of the usual SFA, our model indicates that for a short range potential the main ionisation mechanism does not involve tunneling. Instead, ionisation results from a laser assisted inelastic collision with the residual ion thereby preventing the emission of very slow electrons. The ejection of slow electrons does occur once several excited states are taken into account in the model.\\

This model opens the route to various applications. 
It is ideal to treat the interaction of the hydrogen negative ion considered as a one-active electron system, with an oscillating electric field in the long wave length  limit. Furthermore, this model can be generalized within the mean field approximation to the treatment of the interaction of complex molecules  with attosecond pulses. The study of the interaction of a water molecule with an ultrashort pulse is in progress.

\section{Acknowledgments}
A.G. is "aspirant au Fonds de la Recherche Scientifique (F.R.S-FNRS)". 
Yu.P. thanks the Universit\'e Catholique de Louvain (UCL) for financially supporting several stays at the Institute of Condensed Matter and Nanosciences of the UCL. F.M.F and P.F.O'M gratefully acknowledge the European network COST (Cooperation in Science and Technology) through the Action CM1204 "XUV/X-ray light and fast ions for ultrafast chemistry" (XLIC) for financing several short term scientific missions at UCL. The present research benefited from computational resources made available on the Tier-1 supercomputer of the F\'ed\'eration Wallonie-Bruxelles funded by the R\'egion Wallonne under the grant n$^o$1117545 as well as on the supercomputer Lomonosov from Moscow State University and on the supercomputing facilities of the UCL and the Consortium des Equipements de Calcul Intensif (CECI) en F\'ed\'eration Wallonie-Bruxelles funded by the F.R.S.-FNRS under the convention 2.5020.11. A.G. and Y.P. are grateful to Russian Foundation for Basic Research (RFBR) for the financial support under the grant N14-01-00420-a. B.P. also thanks l' Agence Nationale de la Recherche fran\c{c}aise (ANR) in the context of «Investissements d'avenir» Programme IdEx Bordeaux - LAPHIA (ANR-10-IDEX-03-02).


\begin{thebibliography}{}
\bibitem{remark1} The action of the translation operator $\exp{[i(\vec{r}'-\vec{r})\cdot\vec{p}]}$ on the function $\Psi(\vec{r})$ should provide  the Taylor expansion of 
                                 $\Psi(\vec{r}')$ about its value at $\Psi(\vec{r})$. This is usually  done by Taylor expanding the translation operator itself. But, by proceeding in this way, 
                                 we do not get the correct result because beyond the first order, the action of the operator $\vec{p}$ on $(\vec{r}'-\vec{r})$ introduces extra 
                                 terms in the Taylor expansion of $\Psi(\vec{r}')$.
\bibitem{remark2} It is interesting to mention in this context that moving to the Kramers-Henneberger frame of reference leads also to a field dependent potential.
\end{thebibliography}
\end{document}